\documentclass{article}
\usepackage{spconf,amsmath,graphicx}
\usepackage{multirow}
\usepackage[ruled,linesnumbered]{algorithm2e}
\usepackage{booktabs}
\usepackage{makecell}
\usepackage{array}
\usepackage{amssymb}
\usepackage{tabu}

\title{THE SECOND MULTI-CHANNEL MULTI-PARTY MEETING
TRANSCRIPTION CHALLENGE (M2MeT 2.0): A BENCHMARK FOR SPEAKER-ATTRIBUTED ASR}
\name{\begin{tabular}{c}Yuhao Liang$^{1,2}$, Mohan Shi$^3$, Fan Yu$^2$, Yangze Li$^1$, Shiliang Zhang$^2$, Zhihao Du$^2$, Qian Chen$^2$, \\ Lei Xie$^1$, Yanmin Qian$^4$, Jian Wu, Zhuo Chen, Kong Aik Lee$^{5,6}$, Zhijie Yan$^2$, Hui Bu$^7$\end{tabular}}

\address{
    $^1$Audio, Speech and Language Processing Group (ASLP@NPU), School of Computer Science,\\ Northwestern Polytechnical University, China \\$^2$Speech Lab of DAMO Academy, Alibaba Group, China, \\
    $^3$NERC-SLIP, University of Science and Technology of China (USTC), China \\
    $^4$SpeechLab, Department of Computer Science and Engineering, Shanghai Jiao Tong University, China \\
    $^5$ICT Cluster, Singapore Institute of Technology, Singapore \\
    $^6$Institute for Infocomm Research, A*STAR, Singapore \\
    $^7$Beijing Shell Shell Technology Co., Ltd., Beijing, China \\
    }
\copyrightnotice{979-8-3503-0689-7/23/\$31.00~\copyright2023 IEEE}
\begin{document}
%
\maketitle
\begin{abstract}
\vspace{-0.1cm}
With the success of the first Multi-channel Multi-party Meeting Transcription challenge (M2MeT), the second M2MeT challenge (M2MeT 2.0) held in ASRU2023 particularly aims to tackle the complex task of \emph{speaker-attributed ASR (SA-ASR)}, which directly addresses the practical and challenging problem of ``who spoke what at when" at typical meeting scenario. 
We particularly established two sub-tracks.
The fixed training condition sub-track, where the training data is constrained to predetermined datasets, but participants can use any open-source pre-trained model.
The open training condition sub-track, which allows for the use of all available data and models without limitation.
In addition, we release a new 10-hour test set for challenge ranking. This paper provides an overview of the dataset, track settings, results, and analysis of submitted systems, as a benchmark to show the current state of speaker-attributed ASR.
\end{abstract}
\vspace{-0.3cm}
\section{INTRODUCTION}
\vspace{-0.2cm}
Despite years of research, meeting transcription accuracy still faces significant challenges, including but not limited to overlapping speech, an unknown number of speakers, far-field attenuated speech signals, noise, reverberation, and other factors that can degrade transcription performance. 
The ICASSP2022 Multi-Channel Multi-Party Meeting Transcription (M2MeT) challenge~\cite{Yu2022M2MeT, Yu2022Summary} has played a crucial role in the development of Mandarin meeting transcription technology by addressing the challenge of speech overlap in actual meetings. The challenge consists of two distinct tasks: speaker diarization and multi-speaker automatic speech recognition (ASR). The former involves identifying \emph{who spoke when} in the meeting, while the latter aims to transcribe speech from multiple speakers. In the second M2MeT challenge (M2MeT 2.0), these two tasks are merged into a single speaker-attributed task. 

The M2MeT 2.0 challenge presents two key differences from its predecessor, the first M2MeT. First, the evaluation metric in the first M2MeT challenge was speaker-independent, meaning that transcription could be determined, but the corresponding speaker was not identified. To overcome this limitation and advance current multi-talker ASR systems, the M2MeT 2.0 challenge introduces the \emph{speaker-attributed ASR (SA-ASR)} task. This task not only transcribes the speech but also assigns speaker labels to each transcription. Specifically, we introduce the concatenated minimum permutation character error rate (cpCER) metric to evaluate the performance of the submitted systems. The cpCER is proposed for Mandarin particularly, which is defined similarly to the concatenated minimum permutation word error rate (cpWER)~\cite{watanabe20b_chime}. Second, unlike other related challenges such as Computational Hearing in Multisource Environments (CHIME)~\cite{watanabe20b_chime}, Multimodal Information Based Speech Processing (MISP)~\cite{Chen2022misp}, and M2MeT~\cite{Yu2022M2MeT, Yu2022Summary}, the M2MeT 2.0 challenge offers participants the freedom to utilize any open-source pre-trained model, which is typically prohibited in these challenges. This flexibility aims to explore the feasible industrial application of academic research proposed in previous studies, utilizing various open-source pre-trained models trained on a large amount of data for the SA-ASR task.

The M2MeT 2.0 challenge consists of two sub-tracks:
1) The fixed training condition track, is designed to enable reproducible research in this field by providing a fixed set of training data, open-source pre-trained models, and evaluation criteria.
2) The open training condition track, aims to benchmark state-of-the-art performance in speaker-attributed ASR by allowing participants to use their own data and training techniques.

\vspace{-0.4cm}
\section{RALATED WORKS}
\vspace{-0.3cm}
The SA-ASR task~\cite{DBLP:conf/interspeech/KandaGWMCZY20} involves identifying multiple speakers and transcribing overlapped speech within a single session. One common approach to address this cocktail party challenge is to use speaker diarization to identify the active regions of different speakers. Then, a single-talker ASR system with a speaker separation module can be used to transcribe speech from the known active regions. Alternatively, an end-to-end multi-talker ASR system can be used to transcribe speech and assign speaker labels simultaneously based on corresponding speaker information provided by the diarization system.

Speaker diarization techniques that follow conventional clustering-based approaches usually include two main steps: speaker embedding extraction and clustering. These approaches begin by transforming the input audio stream into a speaker-specific representation, followed by a clustering process like Variational Bayesian HMM clustering (VBx)~\cite{landini2022bayesian} that groups the regions of each speaker into separate clusters. Clustering-based methods typically assign a single speaker label to each frame, making it challenging for them to handle speech overlap.
With the rapid development of deep learning, End-to-End speaker diarization methods like end-to-end neural diarization (EEND)~\cite{DBLP:conf/asru/FujitaKHXNW19} are proposed, leveraging a single neural network to replace the modular cluster-based system.
Inspired by the target speaker extraction~\cite{DBLP:conf/interspeech/WangMWSWHSWJL19, delcroix_tdSpkBeam_ICASSP20, DBLP:conf/icassp/JuRYFLCWXS22, DBLP:conf/slt/JuZRWYXS22}, target speaker voice activity detection (TS-VAD)~\cite{DBLP:conf/interspeech/MedennikovKPKKS20, DBLP:journals/corr/abs-2208-13085, DBLP:conf/icassp/WangQL22} has been proposed, which can estimate the activity level of each speaker in the presence of overlapping speech, providing a promising solution for speaker diarization. 

Single-talker ASR has been extensively investigated in recent years, and various architectures, such as Conformer~\cite{gulati2020conformer}, Branchformer~\cite{DBLP:conf/icml/PengDL022}, and Paraformer~\cite{DBLP:conf/interspeech/GaoZ0Y22}, have been proposed to push the limit of the accuracy of ASR.

Meanwhile, there has been a growing interest in multi-talker ASR, which is designed to transcribe speech containing several speakers. One recent approach, called Speaker-Attributed Transformer(SA-Transformer)~\cite{kanda21b_interspeech}, generates token-level speaker labels during the decoding phase and can directly produce speaker-attributed transcription only using the clustered speaker profile. Moreover, TS-ASR~\cite{DBLP:conf/interspeech/YuDZL022} is also a promising approach that utilizes speaker embedding for target speaker extraction (TSE) and achieves good performance when TSE and ASR are jointly trained. By extracting the speaker embedding, TS-ASR can identify the target speaker's voice and enhance the accuracy of the overall ASR system.

The field of rich transcription with speech overlap has undergone extensive research, with advancements facilitated by numerous challenges and open-source datasets~\cite{Chen2022misp,fiscus2005rich,fiscus2006rich,fiscus2007rich}. Table~\ref{tab:dataset} outlines the primary datasets utilized in this scenario.

WSJ0-2mix~\cite{DBLP:conf/icassp/HersheyCRW16} and Libri2Mix~\cite{cosentino2020librimix} datasets involve mixing pairs of utterances from different speakers at random SNRs, making them primarily used for speech separation tasks where there is full overlap. On the other hand, AMI~\cite{mccowan2005ami}, LibriCSS~\cite{chen2020continuous}, and CHIME-6~\cite{watanabe20b_chime} datasets are recorded in real rooms. However, the AMI dataset's fixed speaker count of four and poor recording quality limit its practical applications. LibriCSS, similar to Libri2Mix, utilizes the Librispeech corpus to produce speech mixes. However, due to the fixed intonation and pace of reading in Librispeech, there remains a disparity between LibriCSS and real meeting scenarios. CHIME-6 dataset is designed for conference or indoor conversation transcription tasks and accounts for overlapped speech. 

Although these datasets significantly contribute to the progress of transcription overlapping speech, they are limited to English. The language barrier poses a challenge in achieving comparable results for non-English languages, such as Mandarin. To overcome this challenge, the AISHELL-4~\cite{fu2021aishell} and AliMeeting~\cite{Yu2022M2MeT} datasets have been developed specifically for Mandarin meeting transcription. AISHELL-4 has a lower overlap ratio, while AliMeeting contains intense discussions. Additionally, AliMeeting records the near-field signal of each participant using a headset microphone, ensuring that only the participant's speech is transcribed.

\vspace{-0.4cm}
\begin{table}[!h]
\caption{Datasets available in the literature in multi-talker speech transcription (OR: overlap rate)}
\setlength{\tabcolsep}{1.0mm}{
\begin{tabular}{@{}ccccc@{}}
\toprule
\multicolumn{1}{l}{Dataset}    & Hours  & \#SPK      & Devices                            & OR (\%)       \\ \midrule
\multicolumn{1}{l}{WSJ0-2mix~\cite{DBLP:conf/icassp/HersheyCRW16}}  & 43     & 129        & Simu                             & Full          \\ \midrule
\multicolumn{1}{l}{Libri2Mix~\cite{cosentino2020librimix}}  & 292    & 1252       & Simu                      & Full          \\ \midrule
\multicolumn{1}{l}{AMI~\cite{mccowan2005ami}}        & 100    & 190        & \makecell[l]{Headset mic, \\ 8-ch mic array} & \textless{}10 \\ \midrule
\multicolumn{1}{l}{LibriCSS~\cite{chen2020continuous}}   & 10     & 40         & 7-ch mic array                     & 0-40          \\ \midrule
\multicolumn{1}{l}{CHIME-6~\cite{watanabe20b_chime}}    & 35     & 26         & 4-ch mic array                     & 40            \\ \midrule
\multicolumn{1}{l}{AISHELL-4~\cite{fu2021aishell}}  & 120    & 61         & 8-ch mic array & 19            \\ \midrule
\multicolumn{1}{l}{AliMeeting~\cite{Yu2022M2MeT}}  & 129    & 481        & \makecell[l]{Headset mic, \\ 8-ch mic array} & 42            \\ \bottomrule
\end{tabular}}
\label{tab:dataset}
\vspace{-0.6cm}
\end{table}

\section{DATASET AND TRACKS}
\vspace{-0.3cm}
AliMeeting~\cite{Yu2022M2MeT, Yu2022Summary}, AISHELL-4~\cite{DBLP:conf/interspeech/FuCLJKCHXWBXDC21}, and CN-Celeb~\cite{DBLP:conf/icassp/FanKLLCCZZCW20} corpus are adopted as our training data, which is the same as the first M2MeT challenge. The AliMeeting dataset is a collection of multi-talker conversation recordings in a meeting setting, comprising a total of 118.75 hours of speech data. The dataset is split into 104.75 hours for training (\emph{Train}), 4 hours for evaluation (\emph{Eval}), and 10 hours for testing (\emph{Test}). The AliMeeting corpus includes both far-field overlapped audios and corresponding near-field audios, which exclusively record and transcribe single-speaker speech. To evaluate the submitted systems, an additional 10 hours of audio data, called \emph{Test-2023}, is incorporated specifically for testing purposes. \emph{Test-2023} comprises 10 sessions that were recorded in 5 different rooms and include 58 speakers. It is crucial to highlight that none of the speakers in \emph{Test-2023} overlap with those in the AliMeeting corpus.
\vspace{-0.3cm}
\begin{figure}[!h]
    \centering
    \includegraphics[scale=0.41]{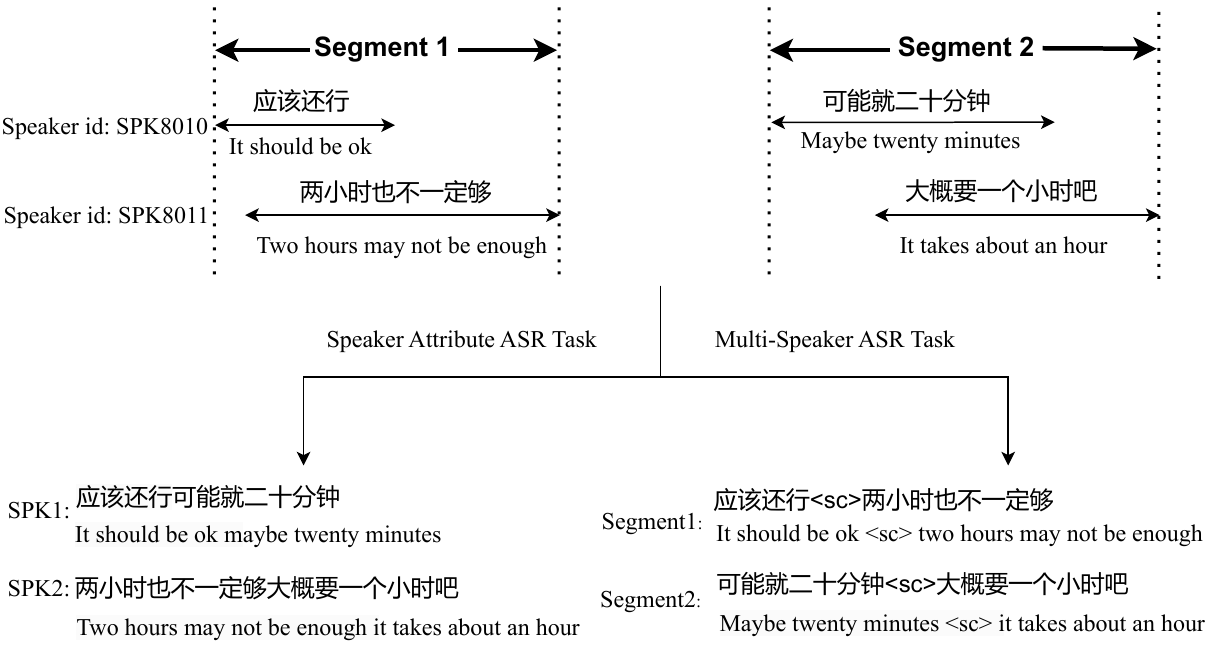}
        \vspace{-0.4cm}
	\caption{
		The main difference between the speaker-attributed ASR task (M2MeT 2.0) and the multi-speaker ASR task (M2MeT).
	}
	\label{fig:example}
\vspace{-0.3cm}
\end{figure}

This challenge introduces the speaker-attributed ASR task which poses a unique challenge of transcribing speech from multiple speakers and assigning a speaker label to the transcription simultaneously. Figure~\ref{fig:example} illustrates the difference between the speaker-attributed ASR task and the multi-speaker ASR task. The speaker-attribution ASR task groups transcriptions from the same speaker together, while the multi-speaker ASR task combines overlapping sentences spoken by different speakers. Sub-track 1 restricts participants from using constrained datasets, while sub-track 2 allows the use of any dataset, including private ones. Any open-source pre-trained models are available for both of these two sub-tracks.
The accuracy of a speaker-attributed ASR system is evaluated using the cpCER. The calculation of cpCER involves three steps.
The first step is to concatenate the reference and hypothesis transcriptions from each speaker in chronological order. This produces a single transcription within one session.
Next, the character error rate (CER) is calculated between the concatenated reference and hypothesis transcriptions, and this process is repeated for all possible speaker permutations.
Finally, the permutation with the lowest CER is selected as the cpCER for that session.

To provide a clear and concise description of the cpCER computation, we illustrate it in Algorithm~\ref{algo:cpcer}. To run the algorithm, it is necessary to provide both the ground truth and hypothesis transcriptions for a given session, which are arranged in chronological order. In cases where the length of $Y$ and $H$ is not equal, we employ padding with blank transcriptions to ensure that both sets have the same length.

\vspace{-0.3cm}
\begin{algorithm}[!ht]
  \SetKwData{Left}{left}\SetKwData{This}{this}\SetKwData{Up}{up}
  \SetKwFunction{Union}{Union}\SetKwFunction{FindCompress}{FindCompress}
  \SetKwInOut{Input}{input}\SetKwInOut{Output}{output}

  \Input{Ground truth: $\left\{Y_1, Y_2, ..., Y_{S}\right\}$, hypothesis of different speakers: $\left\{H_1, H_2, ..., H_{\hat{S}}\right\}$. $S$ is the oracle speaker number and $\hat{S}$ is the predicted speaker number}
  \Output{The $cpCER$ of given session}
  \BlankLine
  $Y\leftarrow \left\{Y_{1}, Y_{2}, ..., Y_{S}\right\}$\;
  $H\leftarrow \left\{H_{1}, H_{2}, ..., H_{\hat{S}}\right\}$\;
  $mindistance\leftarrow INFINITY$\;
  \ForEach {permutation of $H$}{
    $distance\leftarrow 0$\;
    \For {$i\leftarrow 1$ \KwTo $\max(S, \hat{S})$}{
      $distance\leftarrow distance + editdistance(H\left[i\right], Y\left[i\right])$\;
    }
    \If {$distance < mindistance$}{
      $mindistance\leftarrow distance$\;
    }
  }
  $totaltoken\leftarrow\sum_{i=0}^{S}length(Y\left[i\right])$\;
  $cpCER\leftarrow mindistance/totaltoken\times 100\%$
  \caption{Computation of cpCER}\label{algo:cpcer}
\DecMargin{1em}
\end{algorithm}
\setlength{\textfloatsep}{2pt}

\section{SYSTEM DESCRIPTION}
\vspace{-0.2cm}
\subsection{Baseline system}
\begin{figure}[!ht]
    \centering
    \includegraphics[scale=0.7]{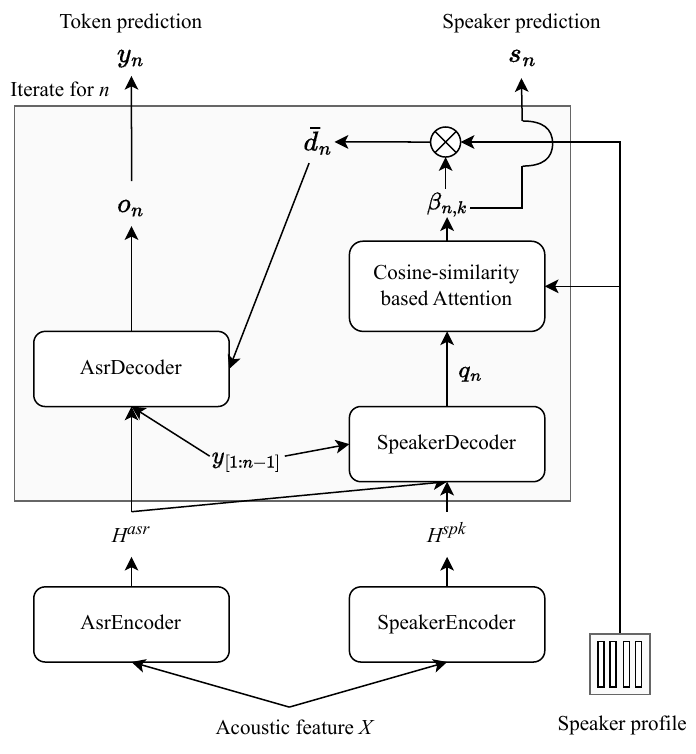}
        \vspace{-0.3cm}
	\caption{
		The SA-Transformer~\cite{kanda21b_interspeech} baseline system uses in M2MeT 2.0 challenge.
	}
	\label{fig:saasr}
\end{figure}

\vspace{-0.1cm}
We release an E2E SA-Transformer baseline built on FunASR~\cite{DBLP:journals/corr/abs-2305-11013} toolkit for easy and reproducible research. The model architecture is illustrated in Figure~\ref{fig:saasr}. It comprises an ASR block and a speaker block to carry out ASR and token-level speaker identification. The ASR block is represented as
\vspace{-0.1cm}
\begin{align}
    H^{asr} &= \text{AsrEncoder}(X), \\
    o_n &= \text{AsrDecoder}(y_{\left[1:n-1\right]}, H^{asr}, \bar{d}_n).
\vspace{-0.1cm}
\end{align}

In the ASR block, the AsrEncoder converts the given acoustic feature $X$ into a series of hidden embeddings $H^{asr}$. The AsrDecoder then produces the output distribution $o_n$ step-by-step. To generate each token, the AsrDecoder takes in the history token $y_{\left[1:n-1\right]}$, the hidden embeddings $H^{asr}$, and the weighted speaker profile $\bar{d}_n$. Compared to the other encoder-decoder-based ASR models, our model differs in its use of the weighted speaker profile $\bar{d}_n$, computed in the speaker block. This profile is employed to bias the transcription towards a specific speaker, which enhances the model's ability to identify individual speakers within a session. The posterior probability of token $i$ at the $n$-th decoding step is formulated as
\vspace{-0.1cm}
\begin{align}
    Pr(y_n = i|y_{\left[1:n-1\right]},s_{\left[1:n\right]}, X, D) = o_{n, i}.
\label{eq:post_asr}
\vspace{-0.1cm}
\end{align}

On the other hand, the speaker block is denoted as
\vspace{-0.1cm}
\begin{align}
    H^{spk} &= \text{SpeakerEncoder}(x), \\
    q_n &= \text{SpeakerDecoder}(y_{\left[1:n-1\right]}, H^{spk}, H^{asr}), \\ 
    \beta_{n,k} &= \frac{\exp(\cos(q_n, d_k))}{ {\textstyle \sum_{j}^{K}} \exp(\cos(q_n, d_j))}, \\
    \bar{d}_n &= \sum_{k=1}^{K} \beta_{n,k}d_k.
\vspace{-0.1cm}
\end{align}
\vspace{-0.1cm}
The speaker encoder takes the input acoustic feature $X$ and produces the speaker embedding $H^{spk}$, which has the same shape as the ASR embedding $H^{asr}$ and represents the speaker's unique characteristics. During each decoding step $n$, the SpeakerDecoder uses $y_{1:n-1}$, $H^{spk}$, and $H^{asr}$ to generate a speaker query $q_n$. Based on this query, a cosine distance-based attention weight $\beta_{n,k}$ is calculated for every profile $d_k$ in $D$. These weights can be viewed as posterior probabilities for the $n$-th token being attributed to the $k$-th speaker, taking into account all previous estimations, as well as relevant information from the input $X$ and the speaker profiles in $D$. The $\beta_{n,k}$ can be represented as 
\begin{align}
    Pr(s_n = k|y_{\left[1:n-1\right]},s_{\left[1:n-1\right]}, X, D) = \beta_{n, k}.
\label{eq:post_spk}
\end{align}
\vspace{-0.1cm}
To extract speaker embeddings and initialize the SpeakerEncoder, we leveraged a pre-trained x-vector extractor from ModelScope trained on CN-Celeb. To train our E2E SA-ASR system, we employed a two-stage training strategy. In the first stage of training the E2E SA-ASR system, we trained a standard Conformer for the ASR task, which was utilized in the second stage to initialize the ASR block. In this second stage, both the ASR and speaker losses were incorporated to fine-tune the model. By using Eqs.~\ref{eq:post_asr} and \ref{eq:post_spk}, the joint posterior of token $Y$ and speaker $S$, optimized to be maximized during training, is represented as
\begin{equation}
\begin{aligned}
    Pr(Y, S|X, D) = \prod_{n=1}^{N} Pr(y_n|y_{\left[1:n-1\right]},s_{\left[1:n\right]}, X, D) \\ \times Pr(s_n|y_{\left[1:n-1\right]},s_{\left[1:n-1\right]}, X, D).
\label{eq:post_joint}
\end{aligned}
\end{equation}

During the training phase, speaker embeddings were extracted from audio solely containing one speaker to generate speaker profiles, utilizing the oracle time stamp. However, during the decoding phase, when the oracle speaker label was absent, we turned to spectral clustering for providing the speaker profile.

\vspace{-0.3cm}
\subsection{Official modular system}
\begin{figure}[!ht]
    \centering
    \includegraphics[scale=0.6]{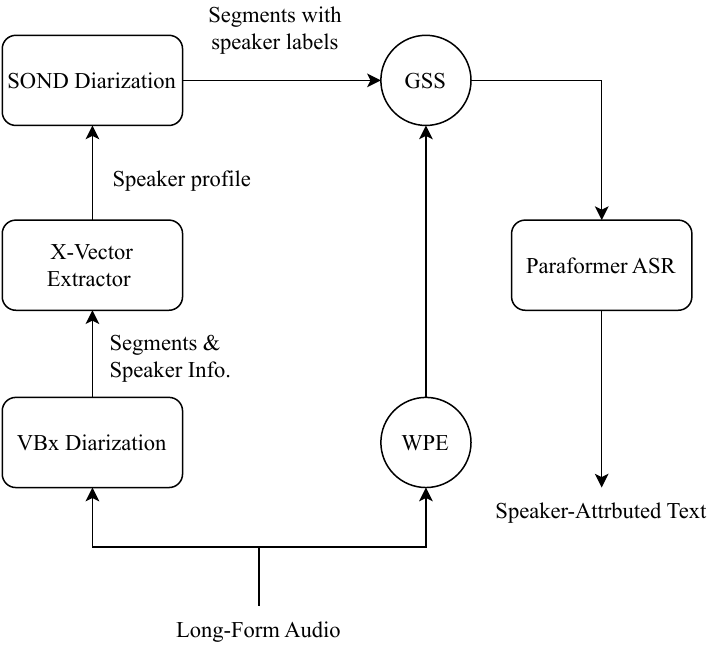}
        \vspace{-0.2cm}
	\caption{
		The official modular system basing on open-source pre-trained models.
	}
	\label{fig:strong_baseline}
\vspace{-0.2cm}
\end{figure}

\vspace{-0.2cm}
In addition to the E2E SA-Transformer baseline, we also develop an official modular system based on pre-trained models that serves as a strong alternative. With the goal of encouraging research and development in the field of speaker-attributed ASR, we are dedicated to releasing this system in the near future. The modular baseline is illustrated in Figure~\ref{fig:strong_baseline}. Our Front-End process leverages two methods to enhance the input audio: weighted prediction error (WPE), a well-known dereverberation technique, and guided source separation (GSS), which utilizes prior segmentation information to assist the separation process. To obtain speaker-labeled segments for GSS, we employ a speaker diarization module. The input long-form audio is first processed by a VBx diarization system to obtain an initial diarization output. Using this output, an x-vector extractor generates a speaker profile for each session from non-overlapping speech segments. The speaker profiles are then utilized by a pre-trained speaker overlap-aware neural diarization (SOND)~\cite{DBLP:conf/emnlp/DuZZY22} model to generate the final diarization output for GSS. Following GSS, we segment the audio based on the diarization output and feed it into a Paraformer for single-talker ASR with speaker labels.

\vspace{-0.3cm}
\section{EXPERIMENTAL SETUP}
\vspace{-0.2cm}
In this section, we describe the settings of the SA-Transformer baseline and the modular system based on the pre-train models. 
\vspace{-0.3cm}
\subsection{SA-Transformer baseline}
\vspace{-0.2cm}
The ASR Encoder is composed of 12 Conformer layers, each equipped with 4-head multi-head attention (MHA). The MHA and feed-forward network (FFN) dimensions are set to 256 and 2048, respectively. The ASR Decoder is made up of 6 transformer layers. The Speaker Encoder is initialized with a pre-trained ResNet x-vector extractor, but with an additional linear layer that transforms the embedding dimension into 256. The Speaker Decoder comprises 3 transformer layers. The speaker loss weight is set to 0.5, while the CTC loss weight is set to 0.3.
\vspace{-0.3cm}
\subsection{Official modular system}
\vspace{-0.2cm}
We have implemented VBx diarization in our modular speaker-attributed system, following the workflow from the first M2MeT challenge baseline. To extract x-vectors, we use the same extractor as in SA-Transformer. When generating speaker profiles, we discard segments shorter than 0.5 seconds. Our system utilizes the pre-trained SOND and Paraformer models, which are both open-sourced and can be accessed on ModelScope. The train set of AliMeeting is utilized to fine-tune the pre-trained paraformer model for 50 epochs with a learning rate of 0.00005.

\vspace{-0.4cm}
\section{RESULTS AND ANLYSIS}
\vspace{-0.3cm}

\begin{table*}[ht!]
\tabcolsep=0.205cm
\caption{Results of top 5 ranking teams in terms of cpCER and their major techniques. Team code O represents the official modular system and Team code B represents the SA-Transformer baseline system}
\newcolumntype{C}[1]{>{\centering\arraybackslash}p{#1}}
\newcolumntype{I}{!{\vrule width 1.4pt}}
\newlength\savewidth
\newcommand\shline{\noalign{\global\savewidth\arrayrulewidth
                           \global\arrayrulewidth 1.4pt}%
                  \hline
                  \noalign{\global\arrayrulewidth\savewidth}}
\scalebox{0.56}{
\begin{tabular}{IcIccIccIcccIccccIccccIcccccIccccccIccIccccI}
\shline
\multirow{2}{*}{\makecell[c]{\\ \\ \\ \\ \\Team \\ Code\\ }} &
  \multicolumn{2}{cI}{Tracks} &
  \multicolumn{2}{cI}{Data} &
  \multicolumn{3}{cI}{Spk-Emb Extractor} &
  \multicolumn{4}{cI}{Diarization} &
  \multicolumn{4}{cI}{Front-End} &
  \multicolumn{5}{cI}{ASR} &
  \multicolumn{6}{cI}{Data Augmentation} &
  \multicolumn{2}{cI}{Post-prosessing} &
  \multicolumn{4}{cI}{cpCER(\%)} \\ \shline
 &
  \multicolumn{1}{c|}{\rotatebox{90}{Sub-track 1}} &
  \multicolumn{1}{cI}{\rotatebox{90}{Sub-track 2}} &
  \multicolumn{1}{c|}{\rotatebox{90}{Constraint Data}} &
  \multicolumn{1}{cI}{\rotatebox{90}{WenetSpeech}} &
  \multicolumn{1}{C{0.54cm}|}{\rotatebox{90}{ResNet-34}} &
  \multicolumn{1}{C{0.54cm}|}{\rotatebox{90}{ResNet-101}} &
  \multicolumn{1}{C{0.54cm}I}{\rotatebox{90}{CAM++}} &
  \multicolumn{1}{c|}{\rotatebox{90}{TS-VAD}} &
  \multicolumn{1}{c|}{\rotatebox{90}{SOND}} &
  \multicolumn{1}{c|}{\rotatebox{90}{Spectral Clustering\qquad}} &
  \multicolumn{1}{cI}{\rotatebox{90}{VBx Clustering}} &
  \multicolumn{1}{c|}{\rotatebox{90}{WPE}} &
  \multicolumn{1}{c|}{\rotatebox{90}{CMGAN}} &
  \multicolumn{1}{c|}{\rotatebox{90}{BeamformIt}} &
  \multicolumn{1}{cI}{\rotatebox{90}{GSS}} &
  \multicolumn{1}{c|}{\rotatebox{90}{U2++}} &
  \multicolumn{1}{c|}{\rotatebox{90}{Paraformer}} &
  \multicolumn{1}{c|}{\rotatebox{90}{MFCCA}} &
  \multicolumn{1}{c|}{\rotatebox{90}{SOT}} &
  \multicolumn{1}{cI}{\rotatebox{90}{SA-Transformer}} &
  \multicolumn{1}{c|}{\rotatebox{90}{Speed Perturb.}} &
  \multicolumn{1}{c|}{\rotatebox{90}{Noise \& Reverb.}} &
  \multicolumn{1}{c|}{\rotatebox{90}{Pitch Shifting}} &
  \multicolumn{1}{c|}{\rotatebox{90}{Data Simulation}} &
  \multicolumn{1}{c|}{\rotatebox{90}{Audio Codec}} &
  \multicolumn{1}{cI}{\rotatebox{90}{SpecAugment}} &
  \multicolumn{1}{C{1cm}|}{\rotatebox{90}{Diarization System Fusion \quad }} &
  \multicolumn{1}{C{1cm}I}{\rotatebox{90}{ASR System Fusion}} &
  \multicolumn{1}{c|}{\makecell[c]{2-Speaker \\ sessions \\ (6) \\ \\ \\ \\ }} &
  \multicolumn{1}{c|}{\makecell[c]{3-Speaker\\ sessions \\ (6) \\ \\ \\ \\ }} &
  \multicolumn{1}{c|}{\makecell[c]{4-Speaker\\ sessions \\ (8) \\ \\ \\ \\ }} &
  \multicolumn{1}{cI}{\makecell[c]{Average\\ \\ \\ \\ \\}} \\ \shline
O &
  \multicolumn{1}{c|}{$\checkmark$} &
  $\checkmark$&
  \multicolumn{1}{c|}{$\checkmark$} &
   &
  \multicolumn{1}{c|}{$\checkmark$} &
  \multicolumn{1}{c|}{$\checkmark$} &
   &
  \multicolumn{1}{c|}{} &
  \multicolumn{1}{c|}{$\checkmark$} &
  \multicolumn{1}{c|}{} &
  $\checkmark$&
  \multicolumn{1}{c|}{$\checkmark$} &
  \multicolumn{1}{c|}{} &
  \multicolumn{1}{c|}{} &
  $\checkmark$ &
  \multicolumn{1}{c|}{} &
  \multicolumn{1}{c|}{$\checkmark$} &
  \multicolumn{1}{c|}{} &
  \multicolumn{1}{c|}{} &
   &
  \multicolumn{1}{c|}{} &
  \multicolumn{1}{c|}{} &
  \multicolumn{1}{c|}{} &
  \multicolumn{1}{c|}{} &
  \multicolumn{1}{c|}{} &
  $\checkmark$ &
  \multicolumn{1}{c|}{} &
   &
  \multicolumn{1}{c|}{9.01} &
  \multicolumn{1}{c|}{7.34} &
  \multicolumn{1}{c|}{10.31} &
  8.84 \\ \hline
X27 &
  \multicolumn{1}{c|}{$\checkmark$} &
  $\checkmark$&
  \multicolumn{1}{c|}{$\checkmark$} &
  &
  \multicolumn{1}{c|}{$\checkmark$} &
  \multicolumn{1}{c|}{} &
   &
  \multicolumn{1}{c|}{$\checkmark$} &
  \multicolumn{1}{c|}{} &
  \multicolumn{1}{c|}{$\checkmark$} &
   &
  \multicolumn{1}{c|}{$\checkmark$} &
  \multicolumn{1}{c|}{} &
  \multicolumn{1}{c|}{$\checkmark$} &
   &
  \multicolumn{1}{c|}{$\checkmark$} &
  \multicolumn{1}{c|}{} &
  \multicolumn{1}{c|}{} &
  \multicolumn{1}{c|}{} &
   &
  \multicolumn{1}{c|}{$\checkmark$} &
  \multicolumn{1}{c|}{} &
  \multicolumn{1}{c|}{} &
  \multicolumn{1}{c|}{$\checkmark$} &
  \multicolumn{1}{c|}{} &
  $\checkmark$ &
  \multicolumn{1}{c|}{$\checkmark$} &
  $\checkmark$ &
  \multicolumn{1}{c|}{11.81} &
  \multicolumn{1}{c|}{9.58} &
  \multicolumn{1}{c|}{12.52} &
  11.27 \\ \hline
M42 &
  \multicolumn{1}{c|}{$\checkmark$} &
  $\checkmark$&
  \multicolumn{1}{c|}{$\checkmark$} &
  &
  \multicolumn{1}{c|}{} &
  \multicolumn{1}{c|}{} &
  $\checkmark$ &
  \multicolumn{1}{c|}{$\checkmark$} &
  \multicolumn{1}{c|}{} &
  \multicolumn{1}{c|}{$\checkmark$} &
   &
  \multicolumn{1}{c|}{} &
  \multicolumn{1}{c|}{$\checkmark$} &
  \multicolumn{1}{c|}{} &
   &
  \multicolumn{1}{c|}{$\checkmark$} &
  \multicolumn{1}{c|}{$\checkmark$} &
  \multicolumn{1}{c|}{} &
  \multicolumn{1}{c|}{} &
   &
  \multicolumn{1}{c|}{$\checkmark$} &
  \multicolumn{1}{c|}{$\checkmark$} &
  \multicolumn{1}{c|}{$\checkmark$} &
  \multicolumn{1}{c|}{$\checkmark$} &
  \multicolumn{1}{c|}{$\checkmark$} &
  $\checkmark$ &
  \multicolumn{1}{c|}{} &
  $\checkmark$ &
  \multicolumn{1}{c|}{18.34} &
  \multicolumn{1}{c|}{16.99} &
  \multicolumn{1}{c|}{20.86} &
  18.64 \\ \hline
C17 &
  \multicolumn{1}{c|}{} &
  $\checkmark$&
  \multicolumn{1}{c|}{$\checkmark$} &
  $\checkmark$&
  \multicolumn{1}{c|}{} &
  \multicolumn{1}{c|}{$\checkmark$} &
   &
  \multicolumn{1}{c|}{} &
  \multicolumn{1}{c|}{} &
  \multicolumn{1}{c|}{} &
  $\checkmark$ &
  \multicolumn{1}{c|}{} &
  \multicolumn{1}{c|}{} &
  \multicolumn{1}{c|}{} &
   &
  \multicolumn{1}{c|}{$\checkmark$} &
  \multicolumn{1}{c|}{} &
  \multicolumn{1}{c|}{} &
  \multicolumn{1}{c|}{} &
   &
  \multicolumn{1}{c|}{} &
  \multicolumn{1}{c|}{} &
  \multicolumn{1}{c|}{} &
  \multicolumn{1}{c|}{} &
  \multicolumn{1}{c|}{} &
  $\checkmark$ &
  \multicolumn{1}{c|}{} &
   &
  \multicolumn{1}{c|}{27.01} &
  \multicolumn{1}{c|}{22.98} &
  \multicolumn{1}{c|}{17.7} &
  22.83 \\ \hline
V29 &
  \multicolumn{1}{c|}{} &
  $\checkmark$&
  \multicolumn{1}{c|}{$\checkmark$} &
  &
  \multicolumn{1}{c|}{$\checkmark$} &
  \multicolumn{1}{c|}{} &
   &
  \multicolumn{1}{c|}{} &
  \multicolumn{1}{c|}{} &
  \multicolumn{1}{c|}{$\checkmark$} &
   &
  \multicolumn{1}{c|}{} &
  \multicolumn{1}{c|}{} &
  \multicolumn{1}{c|}{} &
   &
  \multicolumn{1}{c|}{} &
  \multicolumn{1}{c|}{$\checkmark$} &
  \multicolumn{1}{c|}{} &
  \multicolumn{1}{c|}{} &
   &
  \multicolumn{1}{c|}{} &
  \multicolumn{1}{c|}{} &
  \multicolumn{1}{c|}{} &
  \multicolumn{1}{c|}{} &
  \multicolumn{1}{c|}{} &
  $\checkmark$ &
  \multicolumn{1}{c|}{} &
   &
  \multicolumn{1}{c|}{20.44} &
  \multicolumn{1}{c|}{18.25} &
  \multicolumn{1}{c|}{33.05} &
  23.52 \\ \hline
C31 &
  \multicolumn{1}{c|}{$\checkmark$} &
  $\checkmark$&
  \multicolumn{1}{c|}{$\checkmark$} &
  &
  \multicolumn{1}{c|}{$\checkmark$} &
  \multicolumn{1}{c|}{} &
   &
  \multicolumn{1}{c|}{} &
  \multicolumn{1}{c|}{$\checkmark$} &
  \multicolumn{1}{c|}{} &
   &
  \multicolumn{1}{c|}{} &
  \multicolumn{1}{c|}{} &
  \multicolumn{1}{c|}{} &
   &
  \multicolumn{1}{c|}{} &
  \multicolumn{1}{c|}{} &
  \multicolumn{1}{c|}{$\checkmark$} &
  \multicolumn{1}{c|}{$\checkmark$} &
  $\checkmark$ &
  \multicolumn{1}{c|}{} &
  \multicolumn{1}{c|}{} &
  \multicolumn{1}{c|}{} &
  \multicolumn{1}{c|}{} &
  \multicolumn{1}{c|}{} &
  $\checkmark$ &
  \multicolumn{1}{c|}{} &
   &
  \multicolumn{1}{c|}{21.72} &
  \multicolumn{1}{c|}{19.01} &
  \multicolumn{1}{c|}{34.97} &
  24.82 \\ \hline
B &
  \multicolumn{1}{c|}{$\checkmark$} &
  $\checkmark$&
  \multicolumn{1}{c|}{$\checkmark$} &
  &
  \multicolumn{1}{c|}{$\checkmark$} &
  \multicolumn{1}{c|}{} &
   &
  \multicolumn{1}{c|}{} &
  \multicolumn{1}{c|}{} &
  \multicolumn{1}{c|}{$\checkmark$} &
   &
  \multicolumn{1}{c|}{} &
  \multicolumn{1}{c|}{} &
  \multicolumn{1}{c|}{} &
   &
  \multicolumn{1}{c|}{} &
  \multicolumn{1}{c|}{} &
  \multicolumn{1}{c|}{} &
  \multicolumn{1}{c|}{$\checkmark$} &
  $\checkmark$ &
  \multicolumn{1}{c|}{} &
  \multicolumn{1}{c|}{} &
  \multicolumn{1}{c|}{} &
  \multicolumn{1}{c|}{} &
  \multicolumn{1}{c|}{} &
  $\checkmark$ &
  \multicolumn{1}{c|}{} &
   &
  \multicolumn{1}{c|}{35.95} &
  \multicolumn{1}{c|}{40.49} &
  \multicolumn{1}{c|}{49.4} &
  41.55 \\ \shline
\end{tabular}}
\label{tab:cpcer}
\vspace{-0.6cm}
\end{table*}
In this section, we provide a comprehensive analysis of various systems used in this challenge. The major techniques used and evaluation results of each team is shown in Table~\ref{tab:cpcer}. A total of 30 teams registered for this challenge and 8 of them submitted their results. Almost all teams use constraint data to build their system except C17, so they submit the same result to the two sub-tracks. While we initially released an end-to-end SA-ASR baseline, most participants opted for a modular system approach due to the long training time and limited performance with restricted data for end-to-end approaches. Moreover, with the availability of open-source pre-trained models, developing a modular system is more straightforward and yields satisfactory results. Consequently, we examine different modules, data augmentation techniques, and post-processing methods to ascertain their effectiveness in improving the overall system performance. 
\vspace{-0.3cm}
\subsection{Speaker modules}
\vspace{-0.3cm}
The speaker modules play a crucial role in the speaker-attributed ASR system by providing time-stamp and speaker profiles for subsequent ASR module. The input audio is typically divided into shorter segments, and then a speaker embedding extractor, such as ResNet~\cite{DBLP:conf/cvpr/HeZRS16}, or CAM++~\cite{DBLP:journals/corr/abs-2303-00332}, is employed to encode the audio into an embedding that captures the speaker characteristics. The speaker embedding is used to identify the speaker and corresponding speech segment throughout the ASR process.

The performance of the diarization model is typically measured by the diarization error rate (DER), which is impacted by both speaker activation region prediction and speaker identification accuracy. However, in the context of speaker-attribute ASR tasks, the speaker activation region is not always necessary. As a result, we present the results of speaker counting in Table~\ref{tab:spkcnt}. Teams X27 and M42 both employ TS-VAD with different speaker embedding extractors. By utilizing the clustering result as initialization to further enhance diarization accuracy, team X27 achieves DER of 3.64\% on the \emph{Test} set. Team C17 and V29 opted for spectral clustering and VBx clustering, respectively. Due to the low overlap ratio in the \emph{Test-2023} set, cluster-based diarization methods can also perform well on speaker counting. Team C31 and the official modular system use the SOND model and achieved an accuracy of 80\% and 100\% in speaker counting, respectively. The main difference is that the official modular system uses the VBx to split out the single-talker speech part which can produce more accurate speaker profiles. Benefiting from the accuracy of speaker counting, the official modular system achieves DER of 1.51\% on the \emph{Test-2023} set.

\begin{table}[]
\vspace{-0.2cm}
\tabcolsep=0.32cm
\caption{The speaker counting accuracy (\%) on 20 sessions in \emph{Test-2023} set. $e_{spk}$ and $o_{spk}$ represent the estimated speaker number and oracle speaker number, respectively}
\begin{tabular}{c|ccc}
\toprule
Team code                   & $e_{spk}$\textless{}$o_{spk}$          & $e_{spk}$=$o_{spk}$                   & $e_{spk}$\textgreater{}$o_{spk}$       \\ \midrule
Official       & 0             & 100  & 0               \\ \midrule
X27       & 0             & 90  & 10               \\ \midrule
M42       & 0             & 100 & 0                \\ \midrule
C17       & 10            & 90  & 0                \\ \midrule
V29       & 0             & 95  & 5                \\ \midrule
C31       & 15            & 80  & 5                \\ \midrule
Baseline  & 10            & 50  & 40               \\ \bottomrule
\end{tabular}
\label{tab:spkcnt}
\end{table}
\vspace{-0.4cm}
\subsection{Front-end}
\vspace{-0.2cm}
Out of all the participating teams, only the top 2 teams use the front-end process. The winning team, X27, implements the weighted prediction error (WPE) and weighted delay-and-sum acoustic beamforming (BeamformIt) techniques, which have proven effective in recognizing far-field audio. Team M42 utilizes a Conformer-based Metric GAN (CMGAN)~\cite{DBLP:conf/interspeech/CaoAY22} to separate multi-talker segments into single-talker segments and compared the effectiveness of separate training and joint training strategies. Joint training of the speech separation module and ASR module results in a significant 4.2\% reduction in the character error rate (cpCER) on the \emph{Test-2023} setting (26.60\% $\to$ 22.40\%). The official modular system adopts WPE and GSS. GSS is proven to be an effective method in various challenges when the time stamp can be easily obtained. In the official modular system, it results in a 1.80\% cpCER reduction compared to the process of WPE and BeamformIt (11.98\% $\to$ 10.18\%).

\vspace{-0.4cm}
\subsection{ASR module}
\vspace{-0.2cm}
Multiple ASR models are explored by the participants, including U2++~\cite{DBLP:journals/corr/abs-2106-05642}, Paraformer~\cite{DBLP:conf/interspeech/GaoZ0Y22}, MFCCA~\cite{DBLP:conf/slt/YuZGLDLX22}, and SA-Transformer~\cite{kanda21b_interspeech}. In their experiment, Team M42 utilizes the U2++ and Paraformer models and assessed their performance by comparing the cpCER obtained after joint training with the speaker separation module. The results indicate that Paraformer outperforms U2++ with a lower cpCER of 20.67\% as compared to 22.40\%. Notably, the top four teams processed the audio into single talker segments, which could be effectively handled using a pre-trained standard ASR model.

Team X27~\cite{lyu2023ppmet} successfully transforms a single-talker ASR model into a target speaker model~\cite{DBLP:conf/interspeech/MoriyaSODS22} by incorporating the target speaker's embedding into a fully connected layer and injecting it with an element-wise product between the encoder layers. This approach enabled the model to accurately recognize the target speaker's speech with high precision. Team C31 implements the frame-level diarization with SOT (FD-SOT)~\cite{DBLP:conf/interspeech/YuDZL022} multi-talker ASR system to mitigate the issue of overlap in the ASR module. This multi-talker ASR system combines diarization and ASR results by adhering to the first-in first-out property of SOT. By splitting the ASR result using speaker change symbols and assigning speaker labels using diarization results, FD-SOT ensures one-to-one correspondence. Furthermore, the MFCCA multi-channel solution is adopted as their ASR model, which effectively leverages fine-grained channel-wise information at each time step. It should be noted that MFCCA achieves state-of-the-art level performance with limited data from AliMeeting. However, the conversion of multi-talker ASR transcripts to speaker-attributed ASR transcripts remains a challenging task. 
The official modular system adopts Paraformer as its ASR model. We use the AliMeeting dataset to fine-tune the pre-trained model and achieve the cpCER of 8.84\%, while before fine-tuning, the cpCER is 10.18\%.
\vspace{-0.5cm}
\subsection{Data augmentation and post-processing}
\vspace{-0.2cm}
Data augmentation methods are only adopted by the top 2 teams. One such team, X27, uses the non-overlapping segments in Aishell-4 and CN-Celeb as simulation data to train their TS-VAD model. They have also applied speed perturbation to the training data of their TS-ASR model. Furthermore,  Dover-lap is employed to merge the diarization result from different channels and ROVER is also employed as the final system fusion method, which results in a 0.24\% cpCER reduction on the Test set, reducing it from 17.08\% to 16.84\%.
Team M42 effectively employs a wide range of data augmentation methods, such as speed perturbation, adding noise and reverb, pitch shifting, data simulation, audio codec, and SpecAugment~\cite{DBLP:conf/interspeech/ParkCZCZCL19}. They have also developed U2++ and Paraformer based speaker-attributed systems, and utilize a system fusion method that depends on whether the audio has undergone speech separation. They discover that the conformer model performs better on audio that goes through speech separation while the paraformer model is more effective on audio without speech separation. Therefore, the processed audio is taken as the input of U2++, while the Paraformer takes other inputs to produce the fused hypothesis. This approach results in 2.03\% cpCER reduction on the \emph{Test-2023} set.
\vspace{-0.5cm}
\section{Conclusion}
\vspace{-0.3cm}
This paper provides an overview of the outcomes of the M2MeT 2.0 challenge, with a focus on the techniques used by the top-ranking teams. Given the limited training data and system-building period, leveraging open-source pre-trained models to construct a modular system is an effective approach. For speaker modules, TS-VAD and SOND are potent methods, but accurate speaker profiles are necessary for optimal performance. Front-end processing methods such as WPE, beeamformIt, CMGAN, and GSS are beneficial for transcribing far-field data, although CMGAN may have a negative effect on ASR accuracy for non-overlapped speech. In the ASR module, SOT-based methods underperform due to limited training data, while single-talker ASR models trained on a large amount of data perform well when speech is well-preprocessed in a modular system. Data simulation is less important than that in the first M2MeT challenge, as pre-trained ASR models provide sufficient initialization for fine-tuning with a small amount of data. The best-performing system achieves 8.84\% cpCER given the limited training data of the challenge.

\bibliographystyle{IEEEbib}
\bibliography{strings,refs}

\begin{thebibliography}{10}

\bibitem{Yu2022M2MeT}
Fan Yu, Shiliang Zhang, Yihui Fu, Lei Xie, Siqi Zheng, Zhihao Du, et~al.,
\newblock ``M2{M}e{T}: The {ICASSP} 2022 multi-channel multi-party meeting
  transcription challenge,''
\newblock in {\em Proc. ICASSP}. IEEE, 2022.

\bibitem{Yu2022Summary}
Fan Yu, Shiliang Zhang, Pengcheng Guo, Yihui Fu, Zhihao Du, Siqi Zheng, Lei
  Xie, et~al.,
\newblock ``Summary on the {ICASSP} 2022 multi-channel multi-party meeting
  transcription grand challenge,''
\newblock in {\em Proc. ICASSP}. IEEE, 2022.

\bibitem{watanabe20b_chime}
Shinji Watanabe, Michael Mandel, Jon Barker, et~al.,
\newblock ``{CHiME-6 Challenge: Tackling Multispeaker Speech Recognition for
  Unsegmented Recordings},''
\newblock in {\em Proc. CHiME}, 2020, pp. 1--7.

\bibitem{Chen2022misp}
Hang Chen, Hengshun Zhou, Jun Du, Chin-Hui Lee, Jingdong Chen, Shinji Watanabe,
  Sabato~Marco Siniscalchi, Odette Scharenborg, Di-Yuan Liu, Bao-Cai Yin, Jia
  Pan, Jian-Qing Gao, and Cong Liu,
\newblock ``The first multimodal information based speech processing ({MISP})
  challenge: Data, tasks, baselines and results,''
\newblock in {\em Proc. ICASSP}. IEEE, 2022, pp. 9266--9270.

\bibitem{DBLP:conf/interspeech/KandaGWMCZY20}
Naoyuki Kanda, Yashesh Gaur, Xiaofei Wang, Zhong Meng, Zhuo Chen, Tianyan Zhou,
  and Takuya Yoshioka,
\newblock ``Joint speaker counting, speech recognition, and speaker
  identification for overlapped speech of any number of speakers,''
\newblock in {\em Proc. INTERSPEECH}. 2020, pp. 36--40, {ISCA}.

\bibitem{landini2022bayesian}
Federico Landini, J{\'a}n Profant, Mireia Diez, and Luk{\'a}{\v{s}} Burget,
\newblock ``Bayesian hmm clustering of x-vector sequences (vbx) in speaker
  diarization: theory, implementation and analysis on standard tasks,''
\newblock {\em Proc. CSL}, p. 101254, 2022.

\bibitem{DBLP:conf/asru/FujitaKHXNW19}
Yusuke Fujita, Naoyuki Kanda, Shota Horiguchi, Yawen Xue, Kenji Nagamatsu, and
  Shinji Watanabe,
\newblock ``End-to-end neural speaker diarization with self-attention,''
\newblock in {\em Proc. ASRU}. 2019, pp. 296--303, {IEEE}.

\bibitem{DBLP:conf/interspeech/WangMWSWHSWJL19}
Quan Wang, Hannah Muckenhirn, Kevin~W. Wilson, Prashant Sridhar, Zelin Wu,
  John~R. Hershey, Rif~A. Saurous, Ron~J. Weiss, Ye~Jia, and Ignacio
  Lopez{-}Moreno,
\newblock ``Voicefilter: Targeted voice separation by speaker-conditioned
  spectrogram masking,''
\newblock in {\em Proc. INTERSPEECH}. 2019, pp. 2728--2732, {ISCA}.

\bibitem{delcroix_tdSpkBeam_ICASSP20}
Marc Delcroix, Tsubasa Ochiai, Katerina Zmolikova, Keisuke Kinoshita, Naohiro
  Tawara, Tomohiro Nakatani, and Shoko Araki,
\newblock ``Improving speaker discrimination of target speech extraction with
  time-domain speakerbeam,''
\newblock in {\em Proc. ICASSP}. 2020, pp. 691--695, IEEE.

\bibitem{DBLP:conf/icassp/JuRYFLCWXS22}
Yukai Ju, Wei Rao, Xiaopeng Yan, Yihui Fu, Shubo Lv, Luyao Cheng, Yannan Wang,
  Lei Xie, and Shidong Shang,
\newblock ``{TEA-PSE:} tencent-ethereal-audio-lab personalized speech
  enhancement system for {ICASSP} 2022 {DNS} challenge,''
\newblock in {\em Proc. ICASSP}. 2022, pp. 9291--9295, {IEEE}.

\bibitem{DBLP:conf/slt/JuZRWYXS22}
Yukai Ju, Shimin Zhang, Wei Rao, Yannan Wang, Tao Yu, Lei Xie, and Shidong
  Shang,
\newblock ``{TEA-PSE} 2.0: Sub-band network for real-time personalized speech
  enhancement,''
\newblock in {\em Proc. SLT}. 2022, pp. 472--479, {IEEE}.

\bibitem{DBLP:conf/interspeech/MedennikovKPKKS20}
Ivan Medennikov, Maxim Korenevsky, Tatiana Prisyach, Yuri~Y. Khokhlov, Mariya
  Korenevskaya, Ivan Sorokin, Tatiana Timofeeva, Anton Mitrofanov, Andrei
  Andrusenko, Ivan Podluzhny, Aleksandr Laptev, and Aleksei Romanenko,
\newblock ``Target-speaker voice activity detection: {A} novel approach for
  multi-speaker diarization in a dinner party scenario,''
\newblock in {\em Proc. INTERSPEECH}. 2020, pp. 274--278, {ISCA}.

\bibitem{DBLP:journals/corr/abs-2208-13085}
Dongmei Wang, Xiong Xiao, Naoyuki Kanda, Takuya Yoshioka, and Jian Wu,
\newblock ``Target speaker voice activity detection with transformers and its
  integration with end-to-end neural diarization,''
\newblock in {\em arxiv preprint arxiv:2208.13085}, 2022.

\bibitem{DBLP:conf/icassp/WangQL22}
Weiqing Wang, Xiaoyi Qin, and Ming Li,
\newblock ``Cross-channel attention-based target speaker voice activity
  detection: Experimental results for the m2met challenge,''
\newblock in {\em Proc. ICASSP}. 2022, pp. 9171--9175, {IEEE}.

\bibitem{gulati2020conformer}
Anmol Gulati, James Qin, Chung{-}Cheng Chiu, Niki Parmar, Yu~Zhang, Jiahui Yu,
  Wei Han, Shibo Wang, Zhengdong Zhang, Yonghui Wu, and Ruoming Pang,
\newblock ``Conformer: Convolution-augmented transformer for speech
  recognition,''
\newblock in {\em Proc. INTERSPEECH}. 2020, pp. 5036--5040, ISCA.

\bibitem{DBLP:conf/icml/PengDL022}
Yifan Peng, Siddharth Dalmia, Ian~R. Lane, and Shinji Watanabe,
\newblock ``Branchformer: Parallel mlp-attention architectures to capture local
  and global context for speech recognition and understanding,''
\newblock in {\em Proc. ICML}. 2022, pp. 17627--17643, {PMLR}.

\bibitem{DBLP:conf/interspeech/GaoZ0Y22}
Zhifu Gao, Shiliang Zhang, Ian McLoughlin, and Zhijie Yan,
\newblock ``Paraformer: Fast and accurate parallel transformer for
  non-autoregressive end-to-end speech recognition,''
\newblock in {\em Proc. INTERSPEECH}. 2022, pp. 2063--2067, {ISCA}.

\bibitem{kanda21b_interspeech}
Naoyuki Kanda, Guoli Ye, Yashesh Gaur, Xiaofei Wang, Zhong Meng, Zhuo Chen, and
  Takuya Yoshioka,
\newblock ``{End-to-end speaker-attributed ASR with Transformer},''
\newblock in {\em Proc. INTERSPEECH}. ISCA, 2021, pp. 4413--4417.

\bibitem{DBLP:conf/interspeech/YuDZL022}
Fan Yu, Zhihao Du, Shiliang Zhang, Yuxiao Lin, and Lei Xie,
\newblock ``A comparative study on speaker-attributed automatic speech
  recognition in multi-party meetings,''
\newblock in {\em Proc. INTERSPEECH}. 2022, pp. 560--564, {ISCA}.

\bibitem{fiscus2005rich}
Jonathan~G Fiscus, Nicolas Radde, John~S Garofolo, Audrey Le, Jerome Ajot, and
  Christophe Laprun,
\newblock ``The rich transcription 2005 spring meeting recognition
  evaluation,''
\newblock in {\em Proc. MLMI}. Springer, 2005, pp. 369--389.

\bibitem{fiscus2006rich}
Jonathan~G Fiscus, Jerome Ajot, Martial Michel, and John~S Garofolo,
\newblock ``The rich transcription 2006 spring meeting recognition
  evaluation,''
\newblock in {\em Proc. MLMI}. Springer, 2006, pp. 309--322.

\bibitem{fiscus2007rich}
Jonathan~G Fiscus, Jerome Ajot, and John~S Garofolo,
\newblock ``The rich transcription 2007 meeting recognition evaluation,''
\newblock in {\em Proc. MTPH}. 2007, pp. 373--389, Springer.

\bibitem{DBLP:conf/icassp/HersheyCRW16}
John~R. Hershey, Zhuo Chen, Jonathan~Le Roux, and Shinji Watanabe,
\newblock ``Deep clustering: Discriminative embeddings for segmentation and
  separation,''
\newblock in {\em Proc. ICASSP}. 2016, pp. 31--35, {IEEE}.

\bibitem{cosentino2020librimix}
Joris Cosentino, Manuel Pariente, Samuele Cornell, Antoine Deleforge, and
  Emmanuel Vincent,
\newblock ``Librimix: An open-source dataset for generalizable speech
  separation,''
\newblock in {\em arxiv preprint arxiv:2005.11262}.

\bibitem{mccowan2005ami}
Iain McCowan, Jean Carletta, Wessel Kraaij, Simone Ashby, S~Bourban, M~Flynn,
  et~al.,
\newblock ``The {AMI} meeting corpus,''
\newblock in {\em Proc. ICMT}, 2005, p. 100.

\bibitem{chen2020continuous}
Zhuo Chen, Takuya Yoshioka, Liang Lu, Tianyan Zhou, Zhong Meng, Yi~Luo, Jian
  Wu, Xiong Xiao, and Jinyu Li,
\newblock ``Continuous speech separation: {D}ataset and analysis,''
\newblock in {\em Proc. ICASSP}. IEEE, 2020, pp. 7284--7288.

\bibitem{fu2021aishell}
Yihui Fu, Luyao Cheng, Shubo Lv, Yukai Jv, Yuxiang Kong, Zhuo Chen, Yanxin Hu,
  et~al.,
\newblock ``{AISHELL}-4: {A}n open source dataset for speech enhancement,
  separation, recognition and speaker diarization in conference scenario,''
\newblock in {\em Proc. INTERSPEECH}. ISCA, 2021, pp. 3665--3669.

\bibitem{DBLP:conf/interspeech/FuCLJKCHXWBXDC21}
Yihui Fu, Luyao Cheng, Shubo Lv, Yukai Jv, Yuxiang Kong, Zhuo Chen, Yanxin Hu,
  Lei Xie, Jian Wu, Hui Bu, Xin Xu, Jun Du, and Jingdong Chen,
\newblock ``{AISHELL-4:} an open source dataset for speech enhancement,
  separation, recognition and speaker diarization in conference scenario,''
\newblock in {\em Proc. INTERSPEECH}. 2021, pp. 3665--3669, {ISCA}.

\bibitem{DBLP:conf/icassp/FanKLLCCZZCW20}
Y.~Fan, J.~W. Kang, L.~T. Li, K.~C. Li, H.~L. Chen, S.~T. Cheng, P.~Y. Zhang,
  Z.~Y. Zhou, Y.~Q. Cai, and D.~Wang,
\newblock ``Cn-celeb: {A} challenging chinese speaker recognition dataset,''
\newblock in {\em Proc. ICASSP}. 2020, pp. 7604--7608, {IEEE}.

\bibitem{DBLP:journals/corr/abs-2305-11013}
Zhifu Gao, Zerui Li, Jiaming Wang, Haoneng Luo, Xian Shi, Mengzhe Chen, Yabin
  Li, Lingyun Zuo, Zhihao Du, Zhangyu Xiao, and Shiliang Zhang,
\newblock ``Funasr: {A} fundamental end-to-end speech recognition toolkit,''
\newblock in {\em Proc. INTERSPEECH}. 2023, {ISCA}.

\bibitem{DBLP:conf/emnlp/DuZZY22}
Zhihao Du, Shiliang Zhang, Siqi Zheng, and Zhi{-}Jie Yan,
\newblock ``Speaker overlap-aware neural diarization for multi-party meeting
  analysis,''
\newblock in {\em Proc. EMNLP}, Yoav Goldberg, Zornitsa Kozareva, and Yue
  Zhang, Eds. 2022, pp. 7458--7469, Association for Computational Linguistics.

\bibitem{DBLP:conf/cvpr/HeZRS16}
Kaiming He, Xiangyu Zhang, Shaoqing Ren, and Jian Sun,
\newblock ``Deep residual learning for image recognition,''
\newblock in {\em Proc. CVPR}. 2016, pp. 770--778, {IEEE} Computer Society.

\bibitem{DBLP:journals/corr/abs-2303-00332}
Hui Wang, Siqi Zheng, Yafeng Chen, Luyao Cheng, and Qian Chen,
\newblock ``{CAM++:} {A} fast and efficient network for speaker verification
  using context-aware masking,''
\newblock in {\em arxiv preprint arxiv:2303.00332}, 2023.

\bibitem{DBLP:conf/interspeech/CaoAY22}
Ruizhe Cao, Sherif Abdulatif, and Bin Yang,
\newblock ``{CMGAN:} conformer-based metric {GAN} for speech enhancement,''
\newblock in {\em Proc. INTERSPEECH}, Hanseok Ko and John H.~L. Hansen, Eds.
  2022, pp. 936--940, {ISCA}.

\bibitem{DBLP:journals/corr/abs-2106-05642}
Di~Wu, Binbin Zhang, Chao Yang, Zhendong Peng, Wenjing Xia, Xiaoyu Chen, and
  Xin Lei,
\newblock ``{U2++:} unified two-pass bidirectional end-to-end model for speech
  recognition,''
\newblock in {\em arXiv preprint arXiv:2106.05642}.

\bibitem{DBLP:conf/slt/YuZGLDLX22}
Fan Yu, Shiliang Zhang, Pengcheng Guo, Yuhao Liang, Zhihao Du, Yuxiao Lin, and
  Lei Xie,
\newblock ``{MFCCA}:multi-frame cross-channel attention for multi-speaker {ASR}
  in multi-party meeting scenario,''
\newblock in {\em Proc. SLT}. 2022, pp. 144--151, {IEEE}.

\bibitem{lyu2023ppmet}
Xiang Lyu, Yuhang Cao, Qing Wang, Jingjing Yin, Yuguang Yang, Pengpeng Zou,
  Yanni Hu, and Heng Lu,
\newblock ``Pp-met: a real-world personalized prompt based meeting
  transcription system,''
\newblock in {\em arxiv preprint arXiv:2309.16247}.

\bibitem{DBLP:conf/interspeech/MoriyaSODS22}
Takafumi Moriya, Hiroshi Sato, Tsubasa Ochiai, Marc Delcroix, and Takahiro
  Shinozaki,
\newblock ``Streaming target-speaker {ASR} with neural transducer,''
\newblock in {\em Proc. INTERSPEECH}. 2022, pp. 2673--2677, {ISCA}.

\bibitem{DBLP:conf/interspeech/ParkCZCZCL19}
Daniel~S. Park, William Chan, Yu~Zhang, Chung{-}Cheng Chiu, Barret Zoph,
  Ekin~D. Cubuk, and Quoc~V. Le,
\newblock ``Specaugment: {A} simple data augmentation method for automatic
  speech recognition,''
\newblock in {\em Proc. INTERSPEECH}. 2019, pp. 2613--2617, {ISCA}.

\end{thebibliography}

\end{document}